\documentclass[aps,prl,twocolumn,showpacs,superscriptaddress,amsmath,amssymb]{revtex4-2}
\usepackage{graphicx}
\usepackage{latexsym}
\usepackage{bm} 
\usepackage{color}
\usepackage{epsfig}
\usepackage{multirow}
\usepackage{xcolor}
\usepackage{colortbl}
\usepackage{hhline}
\usepackage{simplewick}
\usepackage{blkarray}
\usepackage{soul}

\usepackage{graphicx}
\usepackage[colorlinks=true,linkcolor=blue,citecolor=blue,urlcolor=blue]{hyperref}
\usepackage{bbold}
\usepackage{gensymb}

\AtBeginDocument{%
    \newwrite\bibnotes
    \def\bibnotesext{Notes.bib}
    \immediate\openout\bibnotes=\jobname\bibnotesext
    \immediate\write\bibnotes{@CONTROL{REVTEX42Control}}
    \immediate\write\bibnotes{@CONTROL{%
    apsrev42Control,author="08",editor="1",pages="0",title="0",year="1"}}
     \if@filesw
     \immediate\write\@auxout{\string\citation{apsrev42Control}}%
    \fi
}%

\def \mbf {\mathbf}

\def \delk {\nabla_{\mbf k}}

\def \curlk {\nabla_{\mbf k}\times}

\newcommand{\bra}[1]{\langle#1|}
\newcommand{\ket}[1]{|#1\rangle}
\newcommand{\braket}[3]{\langle#1|#2|#3\rangle}
\newcommand{\innp}[2]{\langle#1|#2\rangle}

\def \Tr {\mathrm{Tr}}
\newcommand{\comment}[1]{}

\definecolor{mygreen}{rgb}{0, 0.7, 0}

\begin{document}

\title{Geometric semimetals and their simulation in synthetic matter}

\author{Yu-Ping Lin}
\email{yuping.lin@berkeley.edu}
\affiliation{Department of Physics, University of California, Berkeley, California 94720, USA}

\author{Giandomenico Palumbo}
\email{giandomenico.palumbo@gmail.com}
\affiliation{School of Theoretical Physics, Dublin Institute for Advanced Studies, 10 Burlington Road, Dublin 4 D04 C932, Ireland}

\date{\today}

\begin{abstract}
Topological semimetals, such as the Weyl and Dirac semimetals, represent one of the most active research fields in modern condensed matter physics. The peculiar physical properties of these systems mainly originate from their underlying symmetries, emergent relativistic dispersion, and band topology. In this Letter, we present a different class of gapless systems in three dimensions, dubbed \emph{geometric semimetals}. These semimetals are protected by the generalized chiral and rotation symmetries, but are topologically trivial. Nevertheless, we show that their band geometry is nontrivial, as evidenced by the nonzero quantum metric trace with possible quantization. The possible realization in synthetic-matter experiments is also discussed.
\end{abstract}

\maketitle

{\it Introduction.} The search for novel topological semimetals has been a mainstream of modern condensed matter physics. Starting with the Weyl and Dirac semimetals in three dimensions (3D) \cite{armitage18rmp,Wan,Rappe}, which host linear band crossing points (LBCPs) of nondegenerate and doubly degenerate bands, respectively, the exploration has embraced various generalizations of BCs. One important family features the multifold LBCPs with multiple Fermi velocities \cite{lan11prb,manes12prb,wieder16prl,bradlyn16sc,isobe16prb,tang17prl,Hasan,Flicker,rao19n,sanchez19n,schroter19np,takane19prl,lv19prb,boettcher20prl,schroter20sc,lin21prb,lin22prb,Graf2}, which are the higher-spin versions of Weyl and Dirac semimetals. Other families include the BCPs with higher-order dispersions, such as the quadratic BCPs (QBCPs) \cite{moon13prl,kondo15nc,zhu23ax}, and the higher-dimensional BCs in the forms of lines \cite{armitage18rmp,carter12prb,weng15prx,fang15prb,liang16prb,zhao17prl,bzdusek17prb,ahn18prl,lin22prb} or surfaces \cite{liang16prb,bzdusek17prb,turker18prb,wu18prb,lin22prb}. Importantly, the BCs in topological semimetals are associated with suitable topological invariants, which depend mainly on the underlying symmetries and spatial dimensions. These topological invariants can be derived from the Abelian and non-Abelian Berry connections as gauge-invariant quantities. In 3D (5D), the Weyl and chiral multifold semimetals are the momentum-space Dirac (Yang) monopoles with first (second) Chern numbers \cite{chiu16rmp,lin21prb,lin22prb,alpin23prr}. Similarly, 2D and 4D Dirac-like points can act as vortices \cite{lin22prb} and tensor monopoles \cite{palumbo18prl,Palumbo2021}, respectively. Other examples include the Euler numbers of LBCPs \cite{zhao17prl,bouhon,Unal,bouhon2022,jankowski,bouhon2023} and $\mathbb Z_2$ numbers of nodal loops \cite{fang15prb,ahn18prl,wieder,salerno} under combined inversion and time-reversal ($PT$) symmetry, as well as the delicate topology of QBCPs \cite{zhu23ax}. While many topological semimetals have been discovered in quantum materials, recent experiments have demonstrated successful simulations in synthetic matter, including superconducting quantum circuits \cite{tan19prl,zhang2023}, diamond nitrogen-vacancy centers \cite{yu19nsr,chen22s}, and ultracold atoms \cite{Spielman}.

Despite the ubiquitous importance of band topology to the BCs, there exist topologically trivial semimetals which do not carry any topological invariant. Particular examples include the 3D Kane-fermion model \cite{orlita14np,Malcolm,gadge22prb,Teppe} and 2D $\alpha$-$T_3$ model \cite{bercioux11pra,raoux14prl,louvet15prb} under rotation symmetry, where the LBCPs are attached with middle flat bands. On the other hand, a LBCP of $PT$-symmetric nondegenerate bands with $T^2=1$ generally hosts a trivial band topology, since the Berry connection vanishes in the ``real'' band eigenstates (for doubly or quadruply degenerate bands, the LBCPs could be Euler semimetals instead). A natural question then occurs: \emph{Can topologically trivial semimetals receive protection from any symmetry and support nontrivial physical quantities with possible quantization?}

\begin{figure}[b]
\centering
\includegraphics[scale=1]{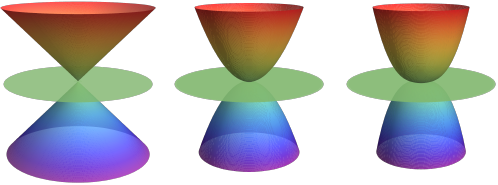}
\caption{\label{fig:colorband} Geometric semimetals. (Left to right) The band structures in the two-spin-sector $N=2$ models with spins $(s_1,s_2)=(0,1)$, $(0,2)$, and $(1,2)$. The dispersive bands are nondegenerate, while the flat-band degeneracies are $2$, $4$, and $6$, respectively.}
\end{figure}

The goal of our {\it Letter} is to provide an affirmative answer to this question. We introduce a general semimetallic model in 3D, where a topologically trivial BCP occurs between multiple dispersive and degenerate middle flat bands (Fig.~\ref{fig:colorband}). Importantly, the semimetals in our model are protected by the generalized chiral and rotation symmetries. Despite trivial band topology, we show that these semimetals host nontrivial structures related to band geometry. The characterization of band geometry lies in the quantum metric \cite{provost80cmp,page87pra,anandan90prl,ma10prb,lin21prb,lin22prb}, which contributes to various observable quantities and physical phenomena \cite{marzari97prb,ma10prb,neupert13prb,Gritsev, roy14prb,peotta15nc,gao15prb,piechon16prb,Thomale,Palumbo2018,Bleu,asteria19np,yu19nsr,tan19prl,Resta,Zhu2,Rastelli,Pozo,Torma,Bernevig,Northe,Graf,ozawa18prb,Ozawa,Mera,Mera2,Wang3,Wang2,kozii21prl,avdoshkin23prb,ahn20prx,Ahn2,hsu23prb,Chen2,Bergholtz,Chen3,Holder,Rossi,Chen4,gianfrate20n,chen22s,Thumin,smith22prr,ledwith23prb,hetenyi23pra}. Note that nontrivial band geometry occurs more ubiquitously than nontrivial band topology: while the latter is strictly related to quantized topological invariants, the former is broadly manifest under nonzero quantum metric. We focus on the quantum metric trace and derive its nontrivial structures with possible quantization \cite{lin21prb,lin22prb}. Given nontrivial band geometry under trivial band topology, we dub our systems \emph{geometric semimetals} to distinguish them from the topological ones. We further show that the variants of our model constitute a broader family, which includes the 3D Kane-fermion model \cite{orlita14np,Malcolm,gadge22prb,Teppe} and 2D $\alpha$-$T_3$ model \cite{bercioux11pra,raoux14prl,louvet15prb} under rotation symmetry. Finally, we discuss the possible experimental realization in synthetic matter.

{\it General model.} We begin by introducing the general model for a class of geometric semimetals in 3D. Our model is a minimal model about a reference point in momentum space. It can be directly simulated in synthetic-matter experiments \cite{tan19prl,zhang2023,yu19nsr,chen22s,Spielman}, or serve as a low-energy $k\cdot p$ theory of suitable lattice models for quantum materials. Without loss of generality, we set the reference point at the momentum-space origin. The central spirit is to couple different spin sectors under the $\text{SU}(2)$ spin-orbit-coupled rotation symmetry. Correspondingly, the relevant ingredients are the total-angular-momentum states
\begin{equation}
\ket{v^{sl}_{jm_j\mbf k}}=\sum_{m_s=-s}^s\sum_{m_l=-l}^l\sqrt{4\pi}\innp{slm_sm_l}{jm_j}Y_{lm_l}(\mbf{\hat k})\ket{sm_s}
\end{equation}
from the additions of spin states $\ket{sm_s}$ and orbital spherical harmonics $Y_{lm_l}(\mbf{\hat k})$ with Clebsch-Gordan coefficients $\innp{slm_sm_l}{jm_j}$. The respective angular momenta and axial components are $s$, $l$, $j$ and $m_{s,l,j}$, which are set as integers in our analysis. Meanwhile, $\mbf{\hat k}$ is the directional unit vector of momentum $\mbf k=(k_1,k_2,k_3)=k\mbf{\hat k}$ with magnitude $k$. For the spin sector $\alpha$ with spin $s_\alpha$, the $j_\alpha=0$ state with $l_\alpha=s_\alpha$
\begin{equation}
\ket{v_{\alpha\mbf k}^1}=\ket{v^{s_\alpha s_\alpha}_{00\mbf k}}   
\end{equation}
is the only rotation symmetric state in the spin-orbit-coupled Hilbert space. This motivates us to consider the $j=0$ projectors between different spin sectors $\alpha\neq\beta$
\begin{equation}
T_{\alpha\beta\mbf k}=k^{s_\alpha}\ket{v_{\alpha\mbf k}^1}\bra{v_{\beta\mbf k}^1}k^{s_\beta},
\end{equation}
which are the rotation symmetric couplings of our interest. We thus define the model with $N$ spin sectors by the Hamiltonian
\begin{equation}
\mathcal H_{\mbf k}=\begin{pmatrix}
0&T_{12\mbf k}&\dots&T_{1N\mbf k}\\
T_{21\mbf k}&0&\ddots&\vdots\\
\vdots&\ddots&\ddots&T_{(N-1)N\mbf k}\\
T_{N1\mbf k}&\dots&T_{N(N-1)\mbf k}&0
\end{pmatrix}
\end{equation}
with $T_{\alpha\beta\mbf k}=T_{\beta\alpha\mbf k}^\dagger$. Note that the diagonal blocks are zero, implying the sole manifestation of the off-diagonal couplings.

\begin{figure}[t]
\centering
\includegraphics[scale=1]{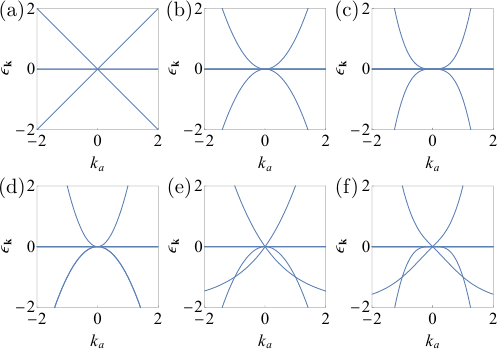}
\caption{\label{fig:band} The band structures in the (a)-(c) $N=2$ models with $(s_1,s_2)=(0,1)$, $(0,2)$, $(1,2)$ and (d)-(f) $N=3$ models with $(s_1,s_2,s_3)=(1,1,1)$, $(0,1,1)$, $(0,1,2)$. All of the dispersive bands are non-degenerate, except for the doubly degenerate negative bands in (d). The flat-band degeneracies are $2$, $4$, $6$, $6$, $4$, $6$ in (a)--(f), respectively. Due to rotation symmetry, the momentum component $k_a$ can be chosen along any direction.}
\end{figure}

Having constructed the general model, we proceed to calculate the dispersion energies $\epsilon_{\mbf k}$ and study the band structure (Fig.~\ref{fig:band}). To secure the BCP at $\epsilon_{\mbf 0}=0$, we assume that the spin-$0$ sector appears at most once. Summing the Hilbert-space dimensions $2s_\alpha+1$ of all spin sectors, the model contains $\sum_{\alpha=1}^N(2s_\alpha+1)$ bands. The first observation is the bundle of middle flat bands with zero dispersion $\epsilon^0_{\mbf k}=0$. These flat bands are composed of the states
\begin{equation}
\ket{u_{\alpha\mbf k}^n}=\dots\oplus\ket{0}\oplus\ket{v_{\alpha\mbf k}^n}\oplus\ket{0}\oplus\dots
\end{equation}
with $n=0$, where the states $\ket{v_{\alpha\mbf k}^0}$ are orthogonal to the $j_\alpha=0$ state $\innp{v_{\alpha\mbf k}^1}{v_{\alpha\mbf k}^0}=0$ in the spin sector $\alpha$. The zero dispersion follows from the projector condition $T_{\alpha\beta\mbf k}\ket{v_{\beta\mbf k}^0}=0$. Since the subspace $\{\ket{v_{\alpha\mbf k}^0}\}$ is $2s_\alpha$-dimensional in each spin sector, the total degeneracy of these flat bands is $\sum_{\alpha=1}^N2s_\alpha$.

There remain $N$ dispersive bands for us to investigate. These bands are composed of the $j_\alpha=0$ states
\begin{equation}
\ket{u_{\mbf k}}=\sum_{\alpha=1}^Nu_{\alpha k}\ket{u_{\alpha\mbf k}^1},
\end{equation}
where the amplitudes $u_{\alpha k}$ are generally $k$-dependent. Projected to the $j_\alpha=0$ basis $\{\ket{u_{\alpha\mbf k}^1}\}_{\alpha=1,2,\dots,N}$, the dispersive Hamiltonian reads
\begin{equation}
\mathcal{\tilde H}_{\mbf k}
=\begin{pmatrix}
0&k^{s_1+s_2}&\dots&k^{s_1+s_N}\\
k^{s_2+s_1}&0&\ddots&\vdots\\
\vdots&\ddots&\ddots&k^{s_{N-1}+s_N}\\
k^{s_N+s_1}&\dots&k^{s_N+s_{N-1}}&0
\end{pmatrix}.
\end{equation}
The diagonalization of this Hamiltonian is complicated in general. Nevertheless, if all effective couplings have the same power $s_\alpha+s_\beta=p$, we can find simple solutions with $k$-independent band eigenstates. This condition applies to all $N=2$ models, as well as the $N>2$ models with the same spin $s_{\alpha=1,2,\dots,N}=p/2$. The corresponding band eigenstates are
\begin{equation}
\begin{aligned}
1\text{ positive band: }&\epsilon^+_{\mbf k}=(N-1)k^p,\\
(N-1)\text{ negative band: }&\epsilon^-_{\mbf k}=-k^p.
\end{aligned}
\end{equation}
The positive band is always non-degenerate, where the eigenstate has a constant amplitude $u_{\alpha\mbf k}=1/\sqrt N$. Meanwhile, the negative band is non-degenerate in the $N=2$ model with $u_{(\alpha=1,2)\mbf k}=(-1)^{\alpha-1}/\sqrt2$.

{\it Symmetry protection.} It is important to understand whether the semimetals are protected by any symmetry. To confirm such a protection, we examine whether a mass term
\begin{equation}
\mathcal H_m=m\begin{pmatrix}
h_{11}&h_{12}&\dots&h_{1N}\\
h_{21}&h_{22}&\ddots&\vdots\\
\vdots&\ddots&\ddots&h_{(N-1)N}\\
h_{N1}&\dots&h_{N(N-1)}&h_{NN}
\end{pmatrix}
\end{equation}
with nonzero mass $m\neq0$ and constant representations $h_{\alpha\beta}=h_{\beta\alpha}^\dagger$ is symmetry-allowed to open a mass gap. Our first target is the chiral symmetry, which plays a crucial role in many semimetals. Interestingly, our general model respects a generalized version of chiral symmetry $\sum_{q=0}^{N-1}U_{S,N}^q\mathcal H_{\mbf k}(U_{S,N}^\dagger)^q=0$ \cite{ni19nm,kempkes19nm}, where $U_{S,N}=\oplus_{\alpha=1}^N\exp[i2\pi(\alpha-1)/N]$ is the generalized chiral unitary operator. In particular, the $N=2$ models respect the standard chiral symmetry $U_S\mathcal H_{\mbf k}U_S^\dagger=-\mathcal H_{\mbf k}$ with $U_S=U_{S,2}$ \cite{chiu16rmp}. The generalized chiral symmetry imposes a strong constraint on the mass term. While nonzero diagonal representations are forbidden $h_{\alpha\alpha}=0$, the off-diagonal ones $h_{\alpha\beta}$ with $\alpha\neq\beta$ are generally allowed. The protection of the semimetals requires additional symmetry to fully reject the remaining terms.

Among various other symmetries, the discrete and continuous rotation symmetries are relevant to some topological semimetals \cite{morimoto14prb,yang15prb,huang22ax}. It turns out that the $\text{SU}(2)$ spin-orbit-coupled rotation symmetry completes the demanded protection in our model. Under the rotation symmetry, the constant representations are only allowed in the identity forms $h_{\alpha\beta}=1$ with $s_\alpha=s_\beta$. If all of the spin sectors carry different spins, $s_\alpha\neq s_\beta$ for $\alpha\neq\beta$, the rotation symmetry forbids all nonzero off-diagonal constant representations $h_{\alpha\beta}=0$. Therefore, the semimetals are protected by the generalized chiral and rotation symmetries. On the other hand, the mass term is allowed if there are different spin sectors $\alpha\neq\beta$ with the same spin $s_\alpha=s_\beta$. In this case, the semimetals lose the protection even if both symmetries are present. Note that BCPs with continuous rotation symmetries can be realized in the synthetic-matter experiments \cite{tan19prl,zhang2023,yu19nsr,chen22s,Spielman}. In the quantum materials with lattice structures, the global symmetries are discrete and the continuous rotation symmetries can be low-energy approximations around some BCPs. These low-energy approximate symmetries, known as the quasisymmetries, have recently been shown to play an important role in gapless topological materials \cite{Guo,Hu}.

One may wonder whether the semimetals receive protection in the tenfold-way classification \cite{chiu16rmp}. Our model respects the time-reversal symmetry $T\mathcal H_{\mbf k}T^{-1}=\mathcal H_{-\mbf k}$ with the symmetry operator $T=U_T\mathcal K$. Here $\mathcal K$ is the complex conjugate operator, and the unitary operator $U_T=\oplus_{\alpha=1}^N\sum_{m_{s_\alpha}=-s_\alpha}^{s_\alpha}(-1)^{s_\alpha+m_{s_\alpha}}\ket{s_\alpha(-m_{s_\alpha})}\bra{s_\alpha m_{s_\alpha}}$ meets the relations of complex conjugate $Y_{lm_l}(\mbf{\hat k})=(-1)^{m_l}Y_{l(-m_l)}^*(\mbf{\hat k})$ and opposite momentum $Y_{lm_l}(\mbf{\hat k})=(-1)^lY_{lm_l}(\mbf{-\hat k})$ for orbital spherical harmonics. Note that the time-reversal operator satisfies $T^2=1$. For the $N>2$ models, the absence of chiral symmetry indicates the AI class. Despite the $\mathbb Z$ classification, the mass term with $h_{\alpha\beta m_{s_\alpha}m_{s_\beta}}=(-1)^{s_\alpha+m_{s_\alpha}+s_\beta+m_{s_\beta}}h_{\alpha\beta (-m_{s_\alpha})(-m_{s_\beta})}^*$ is allowed to gap the semimetal. On the other hand, the $N=2$ models obey the particle-hole symmetry $C\mathcal H_{\mbf k}C^{-1}=-\mathcal H_{-\mbf k}$ with $C=U_C\mathcal{K}$ and $U_C=\oplus_{\alpha=1}^2\sum_{m_{s_\alpha}=-s_\alpha}^{s_\alpha}(-1)^{(\alpha-1)+s_\alpha+m_{s_\alpha}}\ket{s_\alpha(-m_{s_\alpha})}\bra{s_\alpha m_{s_\alpha}}$. The condition $C^2=1$ assigns the models to the BDI class without nontrivial classification. Based on these results, we find no protection of the semimetals in the tenfold-way classification.

Our model also satisfies the reality condition (which is real in the real basis) of $PT$ symmetry $(PT)\mathcal H_{\mbf k}(PT)^{-1}=\mathcal H_{\mbf k}$, which enforces the vanishing of Berry fluxes. Here the symmetry operator $PT=U_{PT}\mathcal K$ with $U_{PT}=\oplus_{\alpha=1}^N\sum_{m_{s_\alpha}=-s_\alpha}^{s_\alpha}(-1)^{m_{s_\alpha}}\ket{s_\alpha(-m_{s_\alpha})}\bra{s_\alpha m_{s_\alpha}}$ meets the complex conjugate relation of orbital spherical harmonics. Under the $PT$ symmetry, the reality condition $h_{\alpha\beta m_{s_\alpha}m_{s_\beta}}=(-1)^{m_{s_\alpha}+m_{s_\beta}}h_{\alpha\beta(-m_{s_\alpha})(-m_{s_\beta})}^*$ does not entirely rule out the mass term. Therefore, the semimetals do not receive protection from the $PT$ symmetry.

{\it Nontrivial band geometry.} We have demonstrated the symmetry protection of the semimetals in our model. A natural question is whether nontrivial band properties related to this protection exist. For the topological semimetals, the robustness is usually linked to certain topological invariants. For example, a chiral semimetal can exhibit Abelian Berry flux $\mbf B_{\mbf k}=\curlk\mbf A_{\mbf k}$ with Berry gauge field (or connection) $\mbf A_{\mbf k}=\braket{u_{\mbf k}}{i\delk}{u_{\mbf k}}$, leading to a finite Chern number $C=(1/2\pi)\oint d\mbf S_{\mbf k}\cdot\mbf B_{\mbf k}$ under the surface integral around the BCP \cite{armitage18rmp}. However, the bands in our model, such as the nondegenerate positive band, exhibit zero Berry flux and trivial band topology under the $PT$ symmetry. This feature can be observed from the spherical harmonics in the band eigenstates, which are the monopole harmonics with zero Berry monopole \cite{wu76npb,wu77prd}.

Despite trivial band topology, our model exhibits nontrivial structures in the band geometry. This nontriviality is characterized by the nonzero quantum metric of the band eigenstates \cite{shankar17chp,lin21prb,lin22prb}
\begin{equation}
g_{ab\mbf k}=\frac{1}{2}\braket{u_{\mbf k}}{\{r_{a\mbf k},r_{b\mbf k}\}}{u_{\mbf k}}.
\end{equation}
Here $a,b$ are the spatial indices, and the position $\mbf r_{\mbf k}=i\delk-\mbf A_{\mbf k}$ is a momentum-space covariant derivative under the Berry gauge field. An important task is to further understand the nontrivial band geometry. Notably, the quantum metric trace
\begin{equation}
\Tr g_{\mbf k}=\braket{u_{\mbf k}}{|\mbf r_{\mbf k}|^2}{u_{\mbf k}}
\end{equation}
represents the momentum-space dual kinetic energy under a position-momentum duality \cite{lin21prb,lin22prb}. With the rotation symmetry, the dual energy is dualized to the free-electron kinetic energy on a spherical shell. This implies the sole contribution from the dynamical angular momentum $\Lambda_{\mbf k}=\mbf r_{\mbf k}\times\mbf k$
\begin{equation}
\Tr g_{\mbf k}=\frac{|\Lambda_{\mbf k}|^2}{k^2}.
\end{equation}
The absence of Berry monopole implies the equivalence between dynamical and orbital angular momenta $\mbf\Lambda_{\mbf k}=\mbf L_{\mbf k}$ \cite{haldane83prl,jainbook,hsiao20prb,lin21prb,lin22prb}. Correspondingly, the dual energy exhibits the quantization $|\Lambda_{\mbf k}|^2=l_\alpha(l_\alpha+1)=s_\alpha(s_\alpha+1)$ in the spin sector $\alpha$. Summing over all spin sectors, the quantum metric trace represents the total dual energy
\begin{equation}
\Tr g_{\mbf k}=\frac{1}{k^2}\sum_{\alpha=1}^N|u_{\alpha k}|^2s_\alpha(s_\alpha+1).
\end{equation}
Note that this result is positively definite, indicating the general presence of nontrivial band geometry in our model. Given nontrivial band geometry under trivial band topology, we name these semimetals as {\it geometric semimetals}.

In general, nontrivial band geometry is not necessarily related to quantized invariants. However, it is worth searching for possible quantization under certain conditions, such as the rotation symmetry. Motivated by the topological invariants, such as the Chern number, we consider the surface integral of quantum metric trace around the BCP \cite{lin21prb}
\begin{equation}
G=\frac{1}{2\pi}\oint d\mbf S_{\mbf k}\cdot\mbf{\hat k}\,\Tr\, g_{\mbf k}.
\end{equation}
This integral generally varies with the choice of surface, since the amplitudes $u_{\alpha\mbf k}$ depend on $k$. Nevertheless, the results become quantized constants in the $N=2$ models
\begin{equation}
G=\sum_{\alpha=1}^2s_\alpha(s_\alpha+1).
\end{equation}
The quantization with respect to angular momentum reflects the protection by rotation symmetry (and the standard chiral symmetry). Therefore, this quantity may serve as a {\it geometric invariant} \cite{lin21prb} for the geometric semimetals. As the simplest examples, we obtain the quantized geometric invariants $G=2$, $6$, and $8$ for $(s_1,s_2)=(0,1)$, $(0,2)$, and $(1,2)$, respectively.

Note that nontrivial band geometry is also present in the degenerate flat bands, which involves the non-Abelian quantum metric \cite{ma10prb,Mera2}. Since the flat-band eigenstates $\ket{u_{\alpha\mbf k}^0}$ are not rotation symmetric, the result is generally anisotropic.

In synthetic-matter experiments, the quantum metric can be extracted directly from the transition rates $\Gamma_a\sim  V^2g_{aa\mbf k}$ under sudden quench or periodic drive at strength $V$ along direction $a$ \cite{ozawa18prb,zhang2023,yu19nsr,chen22s,Spielman}. With this accessibility, the nontriviality and possible quantization can be examined for the quantum metric trace $\Tr g_{\mbf k}$ and geometric invariant $G$, which are physical observables by themselves. On the other hand, the quantum metric trace $\Tr g_{\mbf k}$ is related to the spread of Wannier functions \cite{marzari97prb}, whose gauge-invariant lower bound is supported by the integral of geometric invariant $\Omega_I\sim\int_{\mbf k}\Tr g_{\mbf k}\geq\int_0^{\Lambda_k}dk(2\pi G)$, with a momentum cutoff $\Lambda_k$ around the nodal point \cite{lin21prb}. The corresponding physical observables, including linear injection conductivity \cite{ahn20prx,hsu23prb} and superfluid stiffness \cite{peotta15nc}, are relevant to the quantum-material experiments. Through a direct calculation (see Supplemental Material \cite{supp}), we confirm that the linear injection conductivity indeed receives nontrivial contribution from band geometry in geometric semimetals. Meanwhile, the superfluid stiffness acquires a geometric contribution $\Tr D^S_\text{geom}\gtrsim2\int_0^{\Lambda_k}dk(|\Delta|^2/E)(2\pi G)$ \cite{lin21prb}, where $\Delta$ and $E$ are gap function and quasiparticle energy in the superconductivity.

{\it Variants of general model.} Our analysis has focused on the models with rotation symmetric $j=0$ projectors $T_{\alpha\beta\mbf k}$ between different spin sectors $\alpha\neq\beta$. In fact, the model construction can be generalized to the projectors with $j>0$. For two different spin sectors $\alpha\neq\beta$ with spins $s_\alpha$ and $s_\beta$, we choose the total-angular-momentum states $\ket{v^{s_{\alpha,\beta} l_{\alpha,\beta}}_{j_{\alpha,\beta} m_{j_{\alpha,\beta}}\mbf k}}$ with suitable $j_{\alpha,\beta}=j$ and $l_{\alpha,\beta}$. Although these states break the rotation symmetry, the ``identity'' $T_{\alpha\beta\mbf k}=\sum_{m_j=-j}^jk^{l_\alpha}\ket{v^{s_\alpha l_\alpha}_{jm_j\mbf k}}\bra{v^{s_\beta l_\beta}_{jm_j\mbf k}}k^{l_\beta}$ can serve as a rotation symmetric projector for the variant models. Due to the identity structures of the projectors, the minimal band degeneracy becomes $(2j+1)$. Particular examples include the Kane-fermion model \cite{orlita14np,Malcolm,gadge22prb,Teppe}, where two spin sectors with $(s,l,j)=(1/2,0,1/2)$ and $(3/2,1,1/2)$ are involved. The geometry of doubly degenerate bands is captured by the non-Abelian quantum metric, which does not support a quantized geometric invariant $G$. Another family of variant models involves the dimension reduction to 2D. This is achieved by truncating the elements with $k_3$, leaving only the orbital spherical harmonics $Y_{s(\pm s)}(\mbf{\hat k})$. For example, the 3D model with $(s_1,s_2)=(0,1)$ can be reduced to the $\alpha$-$T_3$ model in 2D \cite{bercioux11pra,raoux14prl,louvet15prb}. Most of the properties can be determined analogously to the 3D models. An important difference is the dual energy in quantum metric trace, which now takes the axial-angular-momentum form $\Lambda_{3\mbf k}^2/k^2=s_\alpha^2/k^2$ in each spin sector $\alpha$ \cite{lin22prb}.

{\it Experimental realization.} We briefly discuss a possible experimental realization of the $N=2$ geometric semimetal with $(s_1,s_2)=(0,1)$.
After a unitary transformation, the Hamiltonian reads
\begin{equation}
\mathcal H_{\mbf k}^{(0,1)}=\frac{1}{\sqrt{2}}\begin{pmatrix}
-k_3 \sqrt{2}& k_1 &k_2&0\\
k_1&0&0&k_1\\
k_2&0&0&k_2\\
0&k_1&k_2&k_3 \sqrt{2}
\end{pmatrix}.
\end{equation}
Interestingly, a similar four-band model \cite{Zhu} has recently been simulated in an experiment \cite{zhang2023}. By employing the superconducting
quantum circuits, a 4D topological semimetal was successfully investigated in a parameter space. The experimental setup involves a square lattice with four transmon qubits and 
four couplers, where the coupling between adjacent qubits depends on the frequency of the couplers. For the simulation of our model $\mathcal H_{\mbf k}^{(0,1)}$, the couplers will need two suitable detuning terms and two independent sinusoidal fast-flux biases with null phases. This setup will support a straightforward mapping between the experimental parameters and the elements of $\mathcal H_{\mbf k}^{(0,1)}$. On the other hand, an alternative setup to simulate our model lies in the ultracold atoms. By employing the atoms with suitable spin degrees of freedom, a four-band model can be simulated, as in a recent realization of the Yang monopole with rubidium-87 \cite{Spielman}.

{\it Conclusion and outlook.} We have demonstrated the existence of geometric semimetals in 3D, which are protected by the generalized chiral and rotation symmetries. Despite trivial band topology, their nontrivial structures are uniquely characterized by the band geometry. These semimetals are thus different from the known topological ones. In future works, we will explore other possible protection symmetries, nonlinear optical responses, interaction effects, and higher-dimensional generalizations. Given the advanced techniques in the quantum-material and synthetic-matter experiments, our theoretical formalism is practically realizable with important experimental implications. Our work embraces the excitements from band geometry, thereby expanding the family of nontrivial semimetals beyond the topological realm. Furthermore, by searching for quantum phases with nontrivial geometry under trivial topology, our work paves a different way into the wide uncharted territory of unconventional quantum matter.

\vspace{0.3cm}

\begin{acknowledgments}
The authors thank Benjamin Wieder for important feedback on this Letter. Y.-P.L. acknowledges fellowship support from the Gordon and Betty Moore Foundation through the Emergent Phenomena in Quantum Systems (EPiQS) program.
\end{acknowledgments}

%%%%%%%%%%

%\appendix

%\section{}
%\label{app:}

\bibliography{ref}

\clearpage
\onecolumngrid

\begin{center}{\large\bf
Supplemental Material for\\``Geometric semimetals and their simulation in synthetic matter"
}\end{center}

\setcounter{secnumdepth}{3}
\setcounter{equation}{0}
\setcounter{figure}{0}
\renewcommand{\theequation}{S\arabic{equation}}
\renewcommand{\thefigure}{S\arabic{figure}}
\newcommand\Scite[1]{[S\citealp{#1}]}
\makeatletter \renewcommand\@biblabel[1]{[S#1]} \makeatother

%\tableofcontents

\section{Linear injection conductivity in geometric semimetals}

Here, for simplicity, we consider our simplest four-band model, namely $\mathcal H_{\mbf k}^{(0,1)}$ with a tilting term. Its momentum-space Hamiltonian is given by
\begin{eqnarray}
	\mathcal H_{\mbf k}^{(0,1)}(v_t,\mu)= k_i T^i + (\mu + v_t k_x) \mathcal{I},
\end{eqnarray}
where $\mathcal{I}$ is the $4 \times 4$ identity matrix, $k_i=\{k_x,k_y,k_z\}$, $\mu$ is the chemical potential, $v_t$ is related to the tilt of the Dirac-like cone along the $x$-direction and
\begin{eqnarray}
T^{x}= \left( \begin{array}{cccc}
0 &  \frac{1}{\sqrt{2}}  & 0 & 0\\
\frac{1}{\sqrt{2}} & 0 & 0 &  \frac{1}{\sqrt{2}}\\
0 & 0 & 0 & 0\\
0 & \frac{1}{\sqrt{2}} & 0 & 0 \end{array} \right), \hspace{0.3cm} 
T^{y}= \left( \begin{array}{cccc}
	0 &  0  & \frac{1}{\sqrt{2}} & 0\\
	0 & 0 & 0 &  0\\
	\frac{1}{\sqrt{2}} & 0 & 0 & \frac{1}{\sqrt{2}}\\
	0 & 0 & \frac{1}{\sqrt{2}} & 0 \end{array} \right), \hspace{0.3cm} 
T^{z}= \left( \begin{array}{cccc}
	-1 &  0  & 0 & 0\\
	0 & 0 & 0 &  0\\
	0 & 0 & 0 & 0\\
	0 & 0 & 0 & 1 \end{array} \right).
\end{eqnarray}
Its spectrum is given by
\begin{eqnarray}
E_1=  (\mu + v_t k_x) - \sqrt{k_x^2+k_y^2+k_z^2}, \hspace{0.3cm}
E_2=  (\mu + v_t k_x), \hspace{0.3cm} \nonumber \\
E_3=  (\mu + v_t k_x) , \hspace{0.3cm}
E_4=  (\mu + v_t k_x) + \sqrt{k_x^2+k_y^2+k_z^2}.
\end{eqnarray}
The corresponding eigenvectors are real and for this reason all the Berry connections are null. However, the Abelian quantum metric is not null. In particular for the eigenvector $V_1$ of the lower band $n=1$ we have
\begin{eqnarray}
	g_{ij}= \partial_i V_1^T  \partial_j V_1,
\end{eqnarray}
and its components are explicitly given by
\begin{eqnarray}
	g_{xy}= -\frac{k_x k_y}{2(k_x^2+k_y^2+k_z^2)^2}, \hspace{0.3cm}
	g_{xx}= \frac{k_y^2+ k_z^2}{2(k_x^2+k_y^2+k_z^2)^2}, \hspace{0.3cm}
	g_{yy}= \frac{k_x^2+ k_z^2}{2(k_x^2+k_y^2+k_z^2)^2}, \hspace{0.3cm} \nonumber \\
	g_{zz}= \frac{k_x^2+ k_y^2}{2(k_x^2+k_y^2+k_z^2)^2}, \hspace{0.3cm}
	g_{xz}= -\frac{k_x  k_z}{2(k_x^2+k_y^2+k_z^2)^2}, \hspace{0.3cm}
	g_{yz}= -\frac{k_y k_z}{2(k_x^2+k_y^2+k_z^2)^2}, \hspace{0.3cm} \nonumber \\
\end{eqnarray}
which do not depend neither of the tilting parameter $v_t$ nor of the chemical potential $\mu$.
We have that
\begin{eqnarray}
	g_{jj} = g_{xx}+g_{yy}+g_{zz}= \frac{1}{|k|^2},
\end{eqnarray}
which is in agreement with our general result in the main text.
In terms of physical observables, we now compute the linear injection in our system. In fact, the quantum metric appears also in several non-linear optical responses. In particular, the linear effect is compatible with space-time inversion $PT$ symmetry, while other non-linear effects such as the circular injection are zero due to the absence of a non-zero Berry curvature.
\noindent Following Ref.~\cite{ahn20prx}, we want now to study the linear injection in our model.
The linear injection is a nonlinear response compatible with $PT$ symmetry and is
related by the change of the electron velocity during the inter-band transition of electrons induced by an oscillating external electric field. 
It is given by
\begin{eqnarray}
\sigma^{c,ab}= - \tau \frac{2 \pi e^3}{\hbar^2} \int \frac{d^3 k}{(2\pi)^3} \sum_{n,m} f_{nm} \Delta^c_{mn} r^b_{nm}r^a_{mn} \delta(E_{mn}-\omega),
\end{eqnarray}
where $\tau$ represents the
relaxation time that saturates the injection current, $\delta$ is the Dirac delta function, $\omega$ is the frequency associated to the external electric field, $f_{nm}=f_n-f_m$, where $f_n$ is the Fermi-Dirac distribution
of the band $n$, $E_{mn}=E_m-E_n$ is the energy difference, $\Delta^c_{mn}=v^c_{mm}-v^c_{nn}$ is the inter-band transition of the velocity with $n$ the occupied bands and $m$ the non-occupied ones. Moreover, $r^a_{mn}=V_m^T i \partial_a V_n$ are the cross gap Berry connections, with $V_n$ the corresponding eigenvectors for the band $n$.
The above expression can be equivalently written in spherical coordinates $(\theta, \phi)$ as follows
\begin{eqnarray}
	\sigma^{c,ab}= - \tau \frac{2 \pi e^3}{\hbar^2} \int_{E_{cv}=\omega} \frac{d\theta d\phi \sin \phi |k|^2}{(2\pi)^3}  (\hat{n} \cdot \hat{c}) g^{n}_{ab},
\end{eqnarray}
because $\Delta_{mn}^c=\partial_c E_{mn}$ such that $\Delta_{mn}^c \delta(E_{mn}-\omega)=\partial_c \Theta(E_{mn}-\omega)$, with $\Theta$ the step function and $e$ the electric charge. Here, $g^{n}_{ab}$ is the quantum metric of the occupied band $n=1$ and the $\hat{n}$ is the normal vector of the surface.
Because the trace of the quantum metric is $1/|k|^2$ as showed previously, the linear injection trace reads
\begin{eqnarray}
	\sigma^{c,aa}= - \tau \frac{e^3}{h^2} \int_{E_{cv}=\omega} d\theta\, d\phi\, \sin \phi\,  (\hat{n} \cdot \hat{c}),
\end{eqnarray}
where the integral depends only of the shape of the Fermi surface, which is completely determined by $v_t$, $\mu$ and $\omega$.
We notice that for the un-tilted case (i.e. $v_t=0$), the surface is a sphere and the above integral is then zero having $\hat{n}=(\sin \phi \cos \theta, \sin \phi \sin \theta, \cos \phi)$.
Thus, tilted cones in geometric semimetals are characterized by a non-zero Fermi-surface-dependent linear injection trace. This is in contract with spin-s Weyl semimetals with tilted Weyl cones, which are also characterized by a circular injection due to the presence of a non-zero Berry curvature. Thus, there exist suitable non-linear physical observables that can naturally distinguish three-dimensional geometric semimetals from three-dimensional topological semimetals.

\end{document}